\documentclass[conference]{IEEEtran}

\ifCLASSINFOpdf
\else
   \usepackage[dvips]{graphicx}
\fi
\usepackage{url}
\usepackage{graphicx}
\usepackage{color}
\usepackage{cite}
\usepackage{amsmath}
\usepackage{amssymb}
\usepackage{subfigure}
\usepackage{algorithm}
\usepackage{algorithmic}
\usepackage{enumitem}
\usepackage{float}

\IEEEoverridecommandlockouts
\begin{document}
\setlength{\topmargin}{-0.7in}

\title{Jointly Sparse Support Recovery via Deep Auto-encoder with Applications in MIMO-based Grant-Free Random Access for mMTC}

\author{
\IEEEauthorblockN{Wanqing~Zhang, Shuaichao~Li, Ying Cui}
\IEEEauthorblockA{Dept. of Electronic Engineering, Shanghai Jiao Tong University, China}
\IEEEauthorblockA{Email: \{wytxz001, lishuaichao, cuiying\}@sjtu.edu.cn}
\thanks{This work was supported in part by the National Key R\&D Program of China under Grant 2018YFB1801102.}
}

\maketitle

\begin{abstract}
In this paper, a data-driven approach is proposed to jointly design the common sensing (measurement) matrix and jointly support recovery method for complex signals, using a standard deep auto-encoder for real numbers. The auto-encoder in the proposed approach includes an encoder that mimics the noisy linear measurement process for jointly sparse signals with a common sensing matrix, and a decoder that approximately performs jointly sparse support recovery based on the empirical covariance matrix of noisy linear measurements. The proposed approach can effectively utilize the feature of common support and properties of sparsity patterns to achieve high recovery accuracy, and has significantly shorter computation time than existing methods. We also study an application example, i.e., device activity detection in Multiple-Input Multiple-Output (MIMO)-based grant-free random access for massive machine type communications (mMTC). The numerical results show that the proposed approach can provide pilot sequences and device activity detection with better detection accuracy and substantially shorter computation time than well-known recovery methods.
\end{abstract}
\begin{IEEEkeywords}
Jointly sparse support recovery, deep learning, auto-encoder, activity detection, grant-free random access.
\end{IEEEkeywords}
\section{Introduction}
Jointly sparse support recovery in Multiple Measurement Vector (MMV) models refers to the estimation of the common support of $M$ jointly sparse $N$-dimensional vectors from $L$ $(\ll N)$ limited noisy linear measurements for each sparse vector based on a common sensing (measurement) matrix. When $M=1$, jointly sparse support recovery reduces down to sparse support recovery in Single Measurement Vector (SMV) models. The jointly sparse support recovery problem (i.e., MMV problem) arises in many applications in communications and signal processing. Two main challenges exist in jointly sparse support recovery. One is to design a common sensing matrix that maximally retains the information on sparsity when reducing signal dimension. The other is to recover the common support with high recovery accuracy and short computation time.

Existing works on jointly sparse support recovery for complex signals consider a given common sensing matrix \cite{tang2010performance,koochakzadeh2018fundamental, 6994860,8437359,8323218,8264818,8846802,liu2018sparse,6158602}. These methods include exhaustive methods \cite{tang2010performance,koochakzadeh2018fundamental}, optimization-based methods such as LASSO \cite{6994860} and Maximum Likelihood (ML) estimation \cite{8437359}, approximate message passing (AMP) \cite{8323218,8264818,8846802,liu2018sparse} and heuristic sparse support recovery algorithms \cite{6158602}. Very few works \cite{koochakzadeh2018fundamental,6994860} investigate the impact of the common sensing matrix on jointly sparse support recovery. It is worth noting that none of \cite{tang2010performance,koochakzadeh2018fundamental, 6994860,8437359,8323218,8264818,8846802,liu2018sparse,6158602} considers the design of the common sensing matrix, or exploits characteristics of sparse patterns for improving recovery accuracy. Hence, the proposed methods in \cite{tang2010performance,koochakzadeh2018fundamental, 6994860,8437359,8323218,8264818,8846802,liu2018sparse,6158602} may not achieve desirable performance for jointly sparse support recovery. In our recent work \cite{8861085}, a data-driven approach is proposed to jointly design the sensing matrix and sparse support recovery method for complex signals in SMV models, using a deep auto-encoder. Our proposed approach achieves substantially higher recovery accuracy with significantly shorter computation time than existing methods when extra structures in sparsity patterns exist. However, directly extending the data-driven approach for SMV models in \cite{8861085} to MMV models cannot explicitly utilize the feature of common support, and hence may not achieve high recovery accuracy for MMV models.

Estimation of a sparse signal itself rather than its support is a closely related topic. In this topic, \cite{sun2016deep,wu2019learning,8262812,shi2019image,8322184} focus on joint design of signal compression and recovery methods for real signals\cite{sun2016deep,wu2019learning,8262812,shi2019image} or complex signals \cite{8322184}, using deep auto-encoders. Note that neither the neural network for complex signals in \cite{8322184} nor direct extensions of the neural networks for real signals to complex signals can achieve linear compression for complex signals. In our recent work \cite{LS2019}, a model-driven approach is proposed to jointly design the sensing matrix and GROUP LASSO-based jointly sparse signal recovery method for complex signals. The proposed Group LASSO-based decoder, which approximates an iterative parallel-coordinate descent algorithm for GROUP LASSO, achieves high recovery accuracy at the cost of computational complexity increase. Note that an effective sensing matrix and recovery method for sparse signal recovery are not necessarily good for support recovery.
\begin{figure*}[t]

\begin{center}
 \resizebox{15.0cm}{!}{\includegraphics{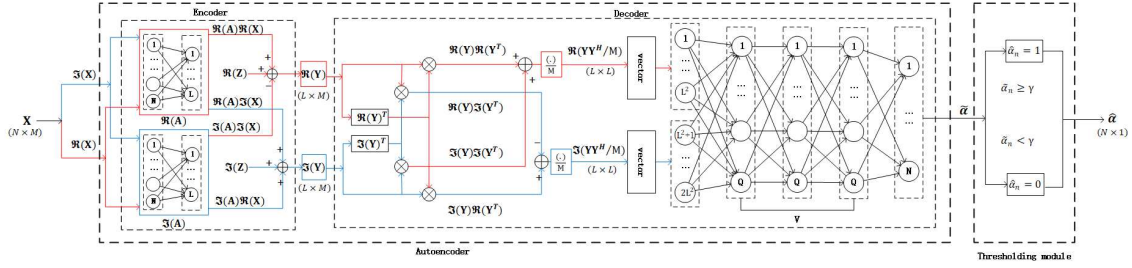}}
  \end{center}
  \vspace*{-0.5cm}
     \caption{Proposed architecture.}
\label{networkmodel1}
\vspace*{-0.5cm}
\end{figure*}

In this paper, our goal is to achieve jointly sparse support recovery for complex signals, with high recovery accuracy and short computation time. To this end, we propose a data-driven approach to jointly design the common sensing matrix and jointly sparse support recovery method for complex signals, using a standard deep auto-encoder for real numbers. The proposed architecture consists of an auto-encoder module and a thresholding module. The auto-encoder includes an encoder that mimics the noisy linear measurement process for jointly sparse signals with a common sensing matrix \cite{LS2019}, and a decoder that approximately performs jointly sparse support recovery based on the empirical covariance matrix of noisy linear measurements. The proposed approach can effectively utilize the feature of common support and properties of sparsity patterns, and is especially useful when it is hard to analytically model the underlying structures of sparsity patterns. In addition, the proposed approach has significantly shorter computation time than model-driven approaches and classic methods, owing to the pure neural network architecture. As an application example, we consider device activity detection in Multiple-Input Multiple-Output (MIMO)-based grant-free random access for massive machine-type communications (mMTC). By numerical results, we demonstrate the substantial gains of the proposed approach over existing methods in terms of both recovery accuracy and computation time.

\textbf{Notation}: We use boldface small letters (e.g., $\mathbf{x}$), boldface capital letters (e.g., $\mathbf{X}$), non-boldface letters (e.g., $x$ or $X$) and calligraphic letters (e.g., $\mathcal{X}$) to represent vectors, matrices, scalar constants and sets, respectively. The notation $X(i,j)$ denotes the $(i,j)$-th element of matrix $\mathbf{X}$, $\mathbf{X}_{i,:}$ represents the $i$-th row of matrix $\mathbf{X}$, $\mathbf{X}_{:,i}$ represents the $i$-th column of matrix $\mathbf{X}$, and $x(i)$ represents the $i$-th element of vector $\mathbf{x}$. Superscript $^H$, superscript $^T$ and superscript $^*$ denote transpose conjugate, transpose and conjugation, respectively. The notation ${\rm vec}(\cdot)$ denotes the column vectorization of a matrix, ${\rm Cov}(\cdot)$ represents the covariance matrix of a random vector, $\odot$ represents the Khatri-Rao product between two matrices, $\mathbb{I}[\cdot]$ denotes the indicator function, and ${\rm Re}(\cdot)$ and ${\rm Im}(\cdot)$ represent the real part and imaginary part, respectively. $\mathbf{0}_{m\times n}$ and $\mathbf{I}_{n\times n}$ represent the $m\times n$ zero matrix and the $n\times n$ identity matrix, respectively. The complex field and real field are denoted by $\mathbb{C}$ and $\mathbb{R}$, respectively.

\section{Jointly Sparse Support Recovery}
\label{sec:format}

The support of a sparse $N$-dimensional complex vector $\mathbf{x} \in \mathbb{C}^N$ is defined as the set of locations of non-zero elements of $\mathbf{x}$, and is denoted by ${\rm supp}(\mathbf{x}) \triangleq \{n\in\mathcal{N}|x(n) \neq 0\}$, where $\mathcal{N} \triangleq \{1,\cdots,N\}$. If the number of non-zero elements of $\mathbf{x}$ is much smaller than its total number of elements, i.e., $|{\rm supp}(\mathbf{x})| \ll N $, $\mathbf{x}$ is sparse. Consider a set of $M$ jointly sparse vectors $\mathbf{x}_m\in \mathbb{C}^N, m \in \mathcal{M} \triangleq \{1,\cdots,M\}$, sharing a common support $\mathcal{S} \triangleq {\rm supp}(\mathbf{x}_m), m \in \mathcal{M}$. Let $\boldsymbol{\alpha}\triangleq (\alpha_n)_{n \in \mathcal{N}}$, where $\alpha_n\triangleq \mathbb{I}[n \in \mathcal{S}]$. That is, $\mathcal{S}=\{n \in \mathcal{N}|\alpha_n=1\}$. For all $m\in \mathcal{M}$, consider $L \ll N$ noisy linear measurements $\mathbf{y}_m\in \mathbb{C}^L$ of $\mathbf{x}_m$, i.e., $\mathbf{y}_m = \mathbf{A}\mathbf{x}_m+\mathbf{z}_m$, where $\mathbf{A}\in \mathbb{C}^{L\times N}$ is the common sensing matrix, and $\mathbf{z}_m\sim\mathcal{CN}(\mathbf{0}_{L\times 1},\sigma^2\mathbf{I}_{L\times L})$ is the additive white Gaussian noise. More compactly, define $\mathbf{X} \in \mathbb{C}^{N\times M}$ with $\mathbf{X}_{:,m} \triangleq \mathbf{x}_m, m\in \mathcal{M}$, $\mathbf{Y}\in \mathbb{C}^{L\times M}$ with $\mathbf{Y}_{:,m} \triangleq \mathbf{y}_m, m\in \mathcal{M}$ and $\mathbf{Z}\in \mathbb{C}^{L\times M}$ with $\mathbf{Z}_{:,m} \triangleq \mathbf{z}_m, m\in \mathcal{M}$. Then, we have:
\vspace*{-0.02cm}
\begin{align}
\mathbf{Y} =\mathbf{A}\mathbf{X}+\mathbf{Z}
\vspace*{-0.2cm}
\end{align}
The jointly sparse support recovery problem, i.e., the MMV problem, aims to identify the common support $\mathcal{S}$ (or $\boldsymbol{\alpha}$) shared by $M$ sparse vectors $\mathbf{x}_m,m \in \mathcal{M}$ (i.e., $\mathbf{X}$) from $M$ noisy linear measurement vectors $\mathbf{y}_m,m \in \mathcal{M}$ (i.e., $\mathbf{Y}$), obtained through a common sensing matrix $\mathbf{A}$ \cite{6158602}. The MMV problem arises in many applications.

As an important application example, we consider device activity detection in MIMO-based grant-free random access, which is recently proposed to support mMTC for IoT \cite{8437359,8323218,8264818,8846802,liu2018sparse}. Consider a single cell with one $M$-antenna base station (BS) and $N$ single-antenna devices. Let $\alpha_n \in \{0,1\}$ represent the active state of device $n$, where $\alpha_n=1$ means that device $n\in\mathcal{N}$ accesses the channel, and $\alpha_n=0$ otherwise. Note that the device activity patterns for IoT traffic are typically sporadic. For all $m \in \mathcal{M}$, let $h_{m}(n)\in \mathbb{C}$ represent the complex channel between the $m$-th antenna at the BS and device $n$, and view $\alpha_nh_{m}(n)$ as $x_m(n)$. Obviously, $\mathbf{x}_{m}\in \mathbb{C}^N,m \in \mathcal{M}$ are sparse with a common support $\mathcal{S}=\{n \in \mathcal{N}|\alpha_n=1\}$. In grant-free random access, each device $n$ has a unique pilot sequence $\mathbf{a}_n \in \mathbb{C}^L$, with $L \ll N$. View $\mathbf{A} \in \mathbb{C}^{L\times N}$ with $\mathbf{A}_{:,n}=\mathbf{a}_n, n\in \mathcal{N}$ as the pilot matrix, which is known at the BS. In the pilot transmission phase, active devices synchronously send their pilot sequences to the BS. Then, $\mathbf{Y}$ in (1) represents the received signal at the BS. The BS conducts device activity detection by estimating $\boldsymbol{\alpha}$ form $\mathbf{Y}$, given knowledge of $\mathbf{A}$, which obviously corresponds to jointly sparse support recovery in MMV models.

\section{Proposed Approach}
\label{approach}
In this section, we propose a data-driven approach, based on the standard auto-encoder structure for real numbers in deep learning, to jointly design the common sensing matrix and the jointly sparse support recovery method for complex signals. As shown in Fig.~\ref{networkmodel1}, the proposed approach consists of an auto-encoder and a thresholding module.

\subsection{Auto-encoder}
First, we illustrate the encoder that mimics the noisy linear measurement process in (1). Note that it has the same structure as the one in our recent work \cite{LS2019}, and is presented here for completeness. To mimic (1) using a standard deep auto-encoder for real numbers, we equivalently express (1) as:
\vspace*{-0.1cm}
\begin{align}
{\rm Re}(\mathbf{Y})={\rm Re}(\mathbf{A}){\rm Re}(\mathbf{X})-{\rm Im}(\mathbf{A}){\rm Im}(\mathbf{X})+{\rm Re}(\mathbf{Z})\\
{\rm Im}(\mathbf{Y})={\rm Im}(\mathbf{A}){\rm Re}(\mathbf{X})+{\rm Re}(\mathbf{A}){\rm Im}(\mathbf{X})+{\rm Im}(\mathbf{Z})
\vspace*{-0.1cm}
\end{align}
Two neural networks, each with two fully-connected layers, are built to implement multiplications with matrices ${\rm Re}(\mathbf{A})\in \mathbb{R}^{L\times N}$ and ${\rm Im}(\mathbf{A})\in \mathbb{R}^{L\times N}$, respectively. For each neural network, there are $N$ neurons and $L$ neurons in the input layer and the output layer, respectively; the weight of the connection from the $n$-th neuron in the input layer to the $l$-th neuron in the output layer corresponds to ${\rm Re}(A(l,n))$ or ${\rm Im}(A(l,n))$; and no activation functions are used in the output layer. The elements of ${\rm Re}(\mathbf{Z})\in \mathbb{R}^{L\times M}$ and ${\rm Im}(\mathbf{Z})\in \mathbb{R}^{L\times M}$ are generated independently according to $\mathcal{N}(0,\frac{\sigma^2}{2})$. As shown in Fig.~\ref{networkmodel1}, when ${\rm Re}(\mathbf{X})\in \mathbb{R}^{N\times M}$ and ${\rm Im}(\mathbf{X})\in \mathbb{R}^{N\times M}$ are input to the encoder, ${\rm Im}(\mathbf{Y})\in \mathbb{R}^{L\times M}$ and ${\rm Re}(\mathbf{Y})\in \mathbb{R}^{L\times M}$ can be easily obtained.

Next, we illustrate the decoder that approximates the jointly sparse support recovery process. Note that one can directly extend the decoder for the SMV problem in \cite{8861085}, without explicitly utilizing the feature of common support. However, the naive approach probably will not provide promising recovery performance for jointly sparse support recovery. This will be seen in Section~\ref{sim}. Motivated by the jointly sparse support recovery method based on the empirical covariance matrix of $M$ linear measurements, i.e., $\mathbf{YY}^H/M$, we propose a novel decoder that can elegantly utilize the feature of common support to effectively improve the performance for jointly sparse support recovery. Specially, by (1), we have
$\mathbf{YY}^H/M = (\mathbf{A}\mathbf{XX}^H\mathbf{A}^H+\mathbf{A}\mathbf{X}\mathbf{Z}^H+\mathbf{Z}\mathbf{X}^H\mathbf{A}^H+\mathbf{ZZ}^H)/M$, which can be equivalently expressed as:
\vspace*{-0.1cm}
\begin{align}
{\rm vec}(\mathbf{YY}^H/M) = \mathbf{A}^*\odot\mathbf{A}\mathbf{r}+{\rm vec}(\mathbf{E_1})+{\rm vec}(\mathbf{E_2})
\vspace*{-0.1cm}
\end{align}
where $\mathbf{r} \in\mathbb{R}^N$ with $r(n) = \frac{\|\mathbf{X}_{n,:}\|_2^2}{M}, n\in\mathcal{N}$, $\mathbf{E}_1\in \mathbb{C}^{L\times L}$ with $E_1(k,l)\triangleq\sum_{i,j\in \mathcal{N},i\neq j}
A(k,i)A^*(l,j)\sum_{m\in \mathcal{M}}x_m(i)x_m^*(j), \\k,l=1,\cdots,L$ and $\mathbf{E}_2=(\mathbf{A}\mathbf{X}\mathbf{Z}^H+\mathbf{Z}\mathbf{X}^H\mathbf{A}^H+\mathbf{ZZ}^H)/M$.
For any given $\mathbf{A}$, if the non-zero elements of $\mathbf{X}$ are i.i.d. random variables with zero mean, then $\mathbf{y}_m,m\in \mathcal{M}$ are i.i.d. random vectors and $\mathbf{YY}^H/M\to{\rm Cov}(\mathbf{y}_m)$, $\mathbf{E}_1\to\mathbf{0}_{L\times L}$ and $\mathbf{E}_2\to\sigma^2\mathbf{I}_{L\times L}$ as $M\to\infty$. Thus, when the non-zero elements of $\mathbf{X}$ are i.i.d. random variables with zero mean and $M\to\infty$, (4) provides linear noiseless measurements of $\mathbf{r}$ with ${\rm supp}(\mathbf{r})={\rm supp}(\mathbf{x}_m),m\in \mathcal{M}$, and hence can be used for jointly sparse support recovery for $\mathbf{X}$. Based on (4), the authors in \cite{6994860} use LASSO for the SMV problem to solve the MMV problem in the case of very large $M$. In Section~\ref{sim}, we shall see that the LASSO-based method in \cite{6994860} does not work well for small $M$ (as $\mathbf{E}_1$ is nonnegligible and $\mathbf{E}_2$ is non-diagonal at small $M$) and has high computational complexity, while the proposed decoder can perfectly resolve these issues.

Now, we introduce the data-driven decoder based on (4), which has a much simpler structure than a model-driven decoder, e.g., the GROUP LASSO-based decoder in \cite{LS2019}. Firstly, as
\vspace*{-0.1cm}
\begin{align}
{\rm Re}(\mathbf{YY}^H)/M=({\rm Re}(\mathbf{Y}){\rm Re}(\mathbf{Y}^T)+{\rm Im}(\mathbf{Y}){\rm Im}(\mathbf{Y}^T))/M\\
{\rm Im}(\mathbf{YY}^H)/M=({\rm Im}(\mathbf{Y}){\rm Re}(\mathbf{Y}^T)-{\rm Re}(\mathbf{Y}){\rm Im}(\mathbf{Y}^T))/M
\vspace*{-0.1cm}
\end{align}
we can obtain ${\rm Re}(\mathbf{YY}^H)/M$ and ${\rm Im}(\mathbf{YY}^H)/M$ based on the output of the encoder ${\rm Im}(\mathbf{Y})$ and ${\rm Re}(\mathbf{Y})$, as shown in Fig.~\ref{networkmodel1}. Then, a fully-connected neural network with $V+2$ layers is built to approximate the jointly sparse support recovery process based on (4), where $V$ is a natural number properly chosen according to the size of the MMV problem. Especially, it includes one input layer, one output layer and $V$ hidden layers. The input layer has $2L^2$ neurons with ${\rm vec}({\rm Re}(\mathbf{YY}^H)/M)$ as the input of the first $L^2$ neurons and ${\rm vec}({\rm Im}(\mathbf{YY}^H)/M)$ as the input of the last $L^2$ neurons. In each of the $V$ hidden layers, there are $Q$ neurons and the rectified linear unit (ReLU) is chosen as the activation function. The output layer has $N$ neurons and the Sigmoid function is chosen as the activation function for producing output $\tilde{\boldsymbol{\alpha}}\in (0,1)^N$ which is used to estimate $\boldsymbol{\alpha}$.

Then, we introduce the training procedure for the proposed approach for jointly sparse support recovery. Choose $U$ training samples $(\mathbf{X}^{[u]}, \boldsymbol{\alpha}^{[u]}),u=1,\cdots,U$. Let $\tilde{\boldsymbol{\alpha}}^{[u]}$ represent the output of the neural network corresponding to input $\mathbf{X}^{[u]}$. To measure the distance between $\boldsymbol{\alpha}^{[u]}$ and $\tilde{\boldsymbol{\alpha}}^{[u]}$, as in \cite{8861085}, the binary cross-entropy loss function which is given by (\ref{equ}), as shown at the top of the next page, is adopted.
\begin{figure*}[htb]
\vspace*{-0.25cm}
\begin{align}
\label{equ}
{\rm Loss}((\boldsymbol{\alpha}^{[u]},\tilde{\boldsymbol{\alpha}}^{[u]})_{u=1,\cdots,U})=\frac{-1}{NU}\sum_{u=1}^{U}\sum_{n=1}^{N}
\big(\alpha(n)^{[u]}\log(\tilde{\alpha}(n)^{[u]})+(1-\alpha(n)^{[u]})\log(1-\tilde{\alpha}(n)^{[u]})\big)
\end{align}
\vspace*{-0.35cm}
\end{figure*}
The ADAM algorithm is used to train the auto-encoder. After training, we obtain the design of the common sensing matrix $\mathbf{A}$ via extracting the weights of the encoder, and directly use the decoder for jointly sparse support recovery together with the obtained common sensing matrix.
\begin{figure*}[t]
\vspace*{-0.25cm}
\hrulefill{
\begin{center}
 \subfigure[\scriptsize{Error rate versus $L/N$ at $p=0.1$, $M=4$, $p_1/p_2=3$, $G=50$.}]
 {\resizebox{5.2cm}{!}{\includegraphics{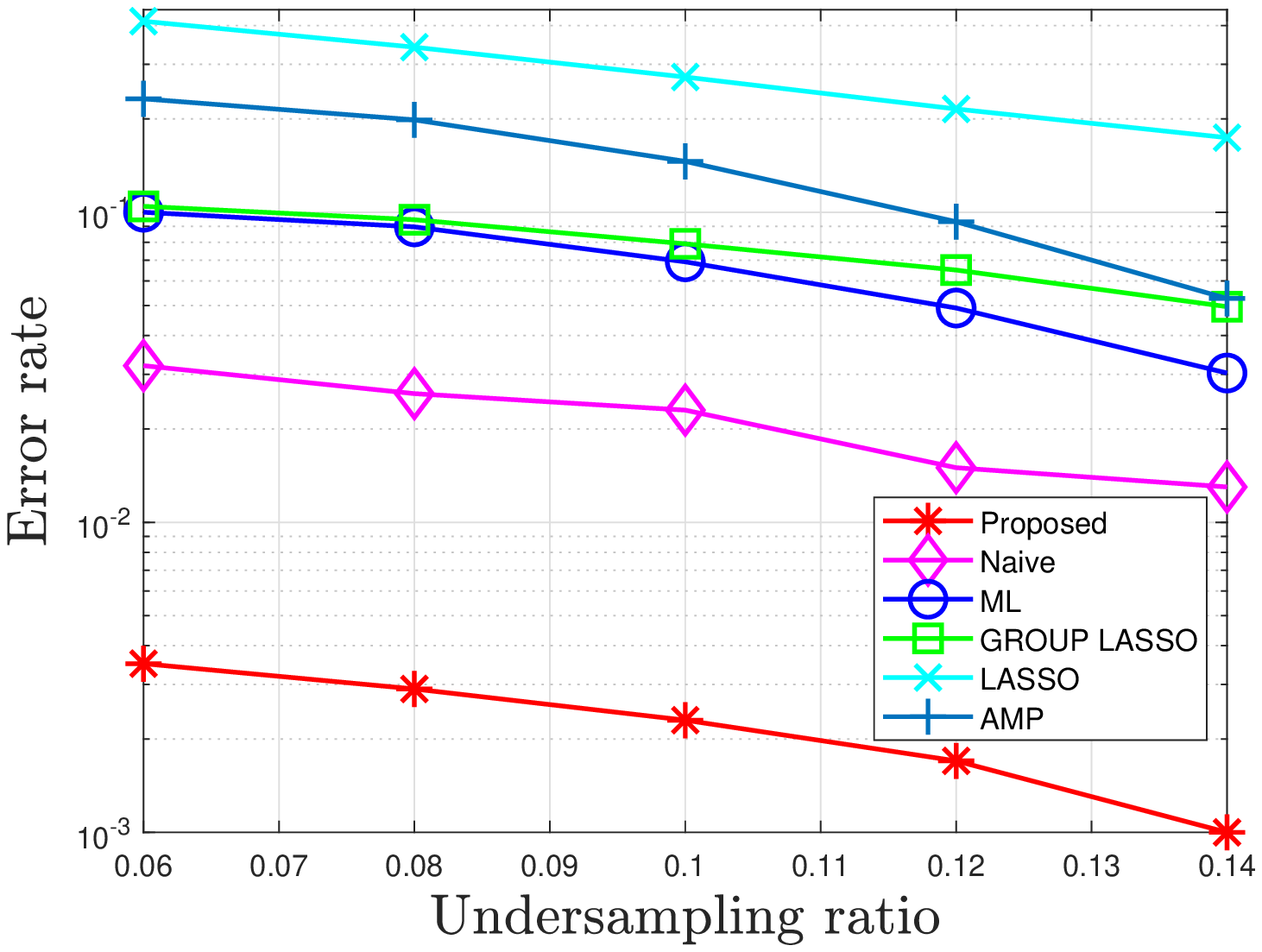}}}
\quad
 \subfigure[\scriptsize{Error rate versus $p$ at $L/N=0.14$, $M=4$, $p_1/p_2=3$, $G=50$.}]
 {\resizebox{5.2cm}{!}{\includegraphics{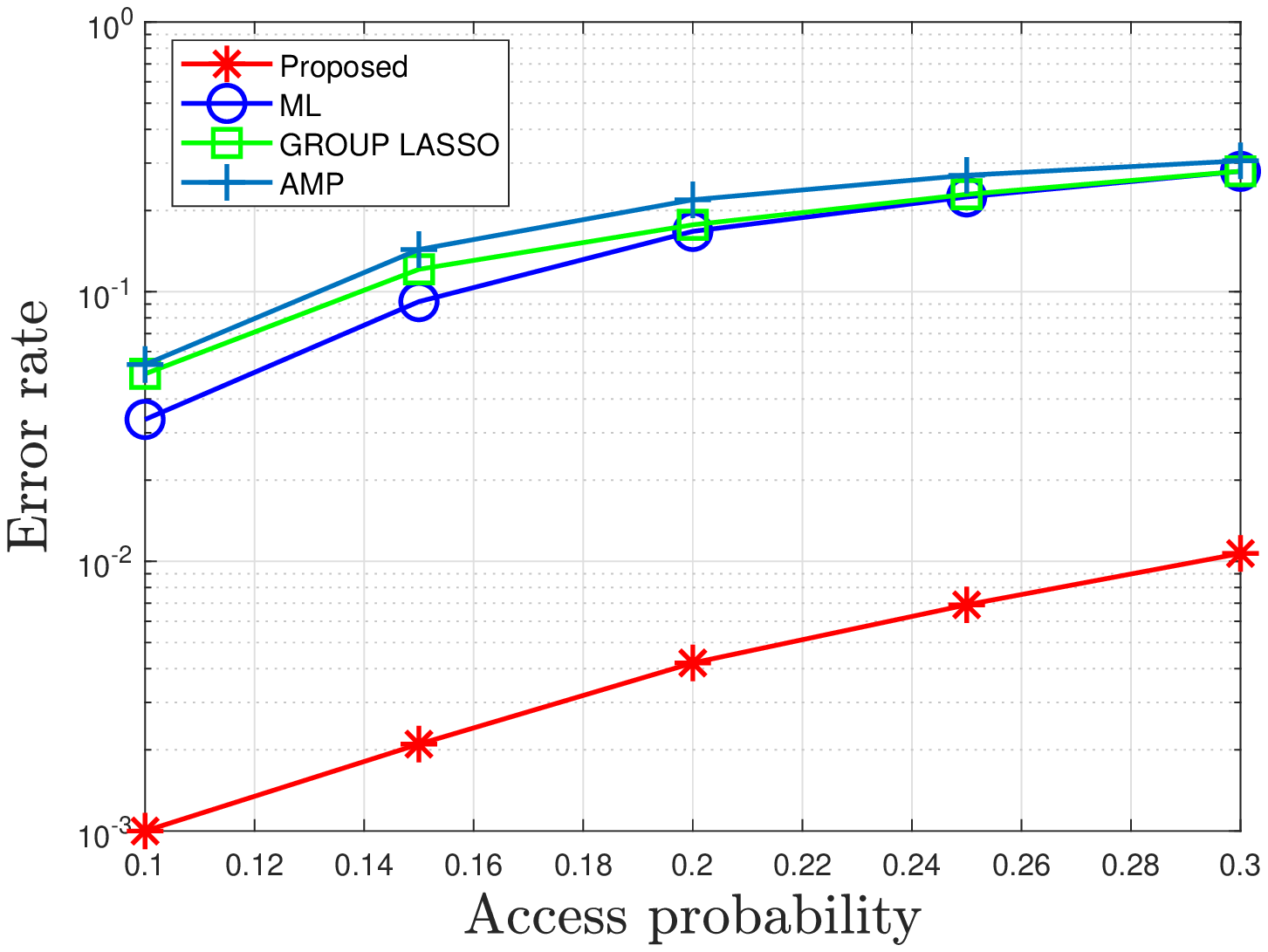}}}
\quad
 \subfigure[\scriptsize{Error rate versus $M$ at $L/N=0.14$, $p=0.1$, $p_1/p_2=3$, $G=50$.}]
 {\resizebox{5.2cm}{!}{\includegraphics{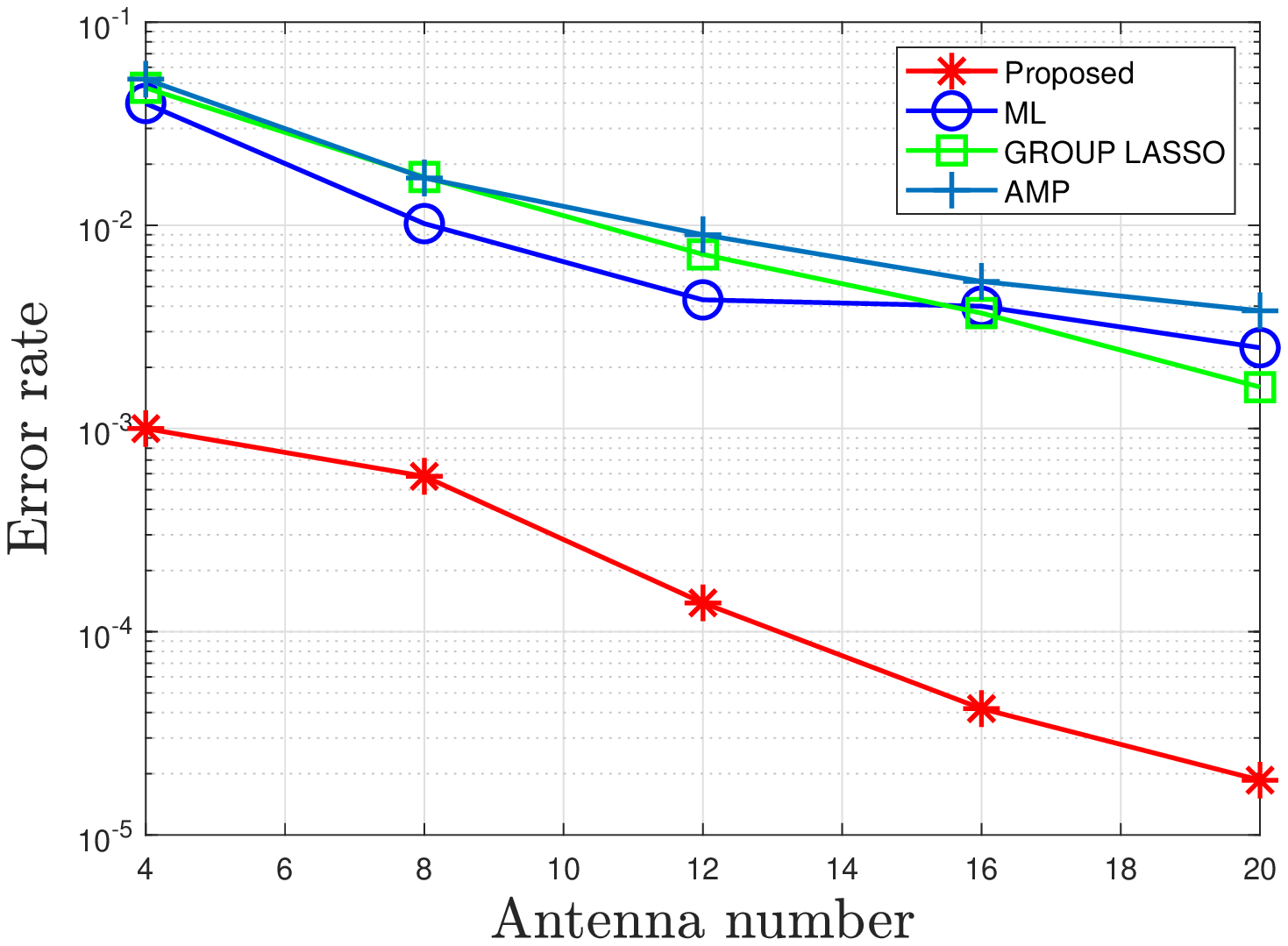}}}
\quad
 \subfigure[\scriptsize{Error rate versus $p_1/p_2$ at $L/N=0.14$, $M=4$, $p=0.1$, $G=50$.}]
 {\resizebox{5.2cm}{!}{\includegraphics{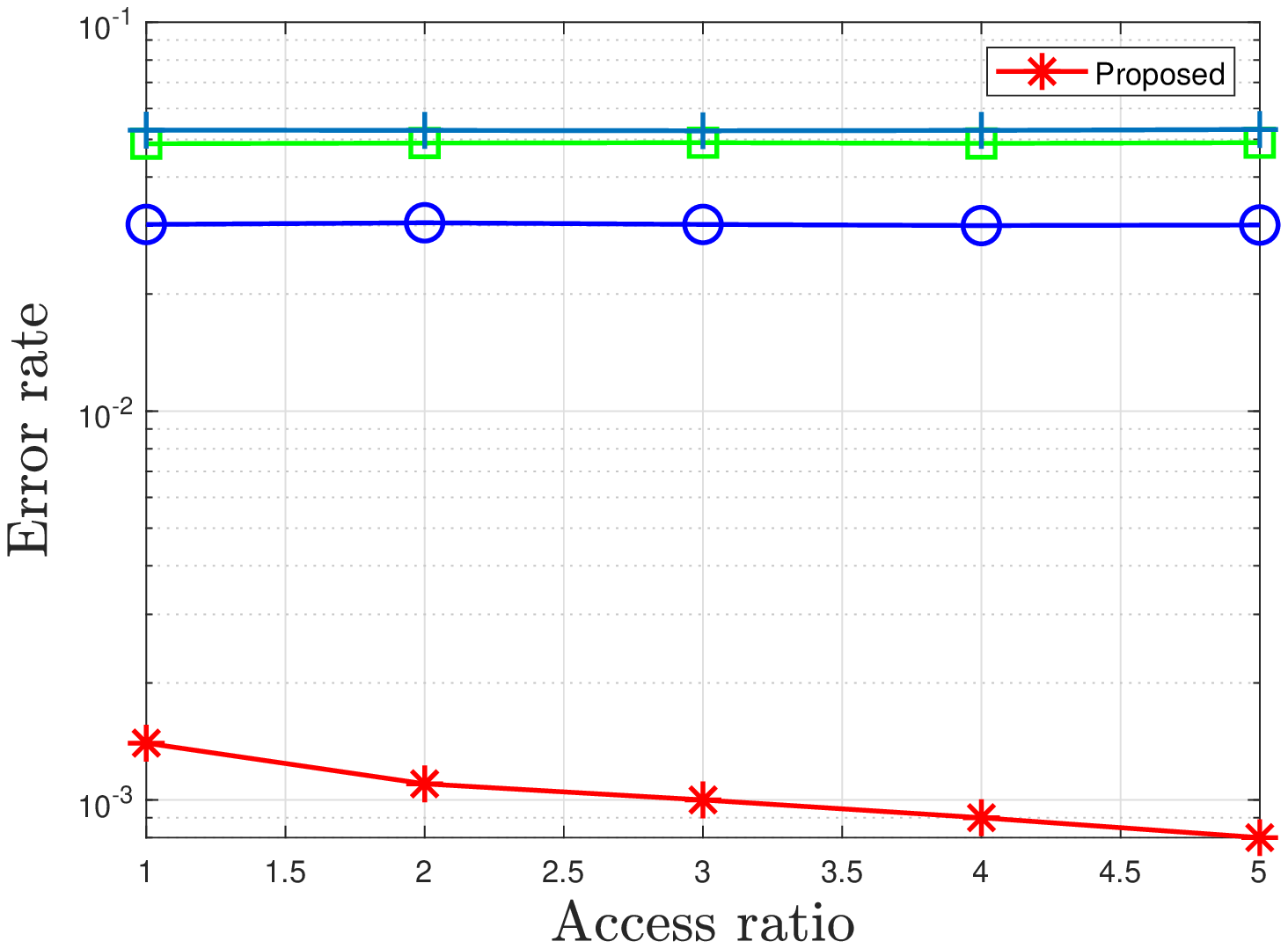}}}
\quad
  \subfigure[\scriptsize{Error rate versus $G$ at $L/N=0.14$, $M=4$, $p=0.1$, $p_1/p_2=3$.}]
 {\resizebox{5.2cm}{!}{\includegraphics{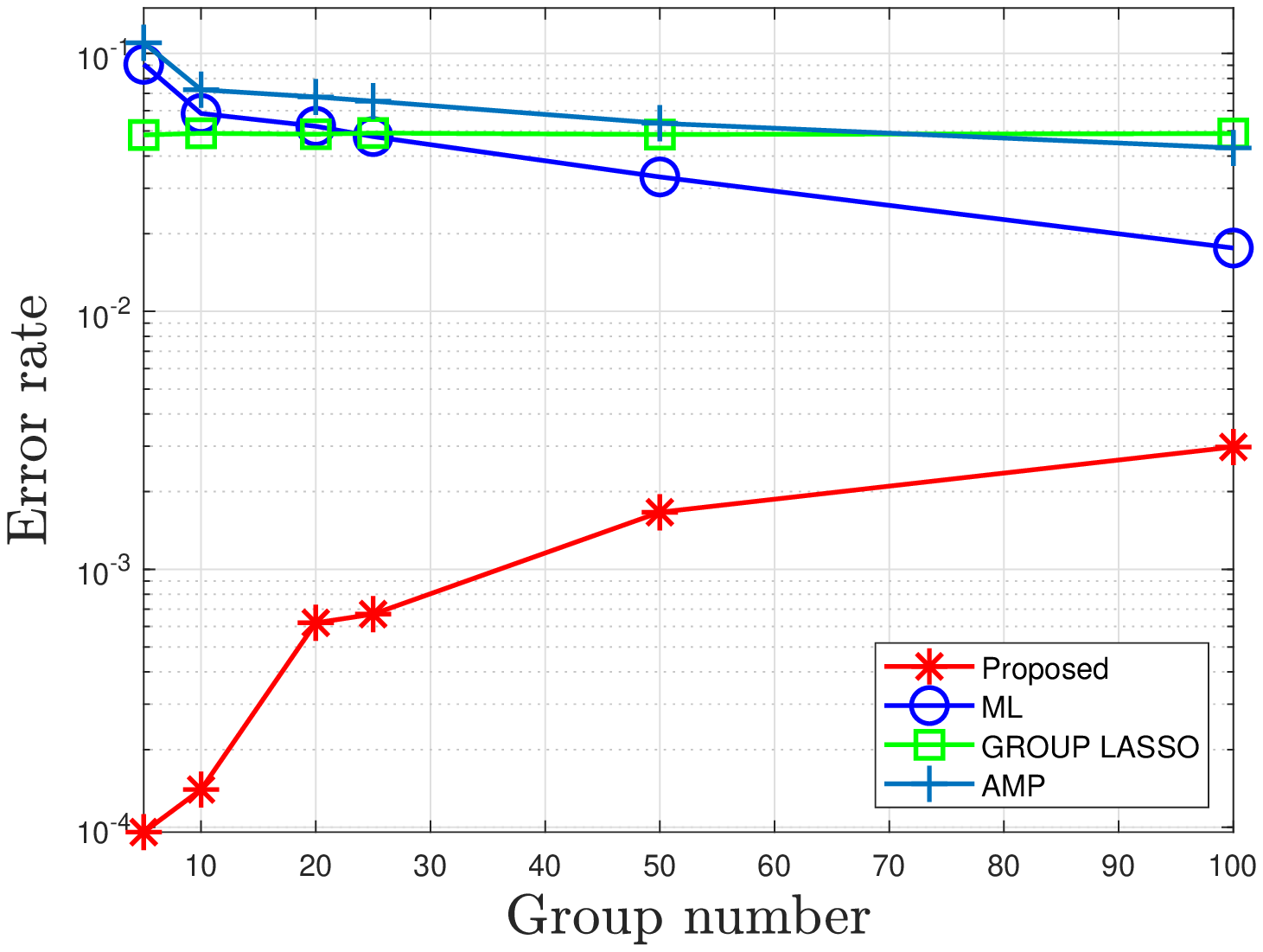}}}
  \end{center}}
  \vspace*{-0.5cm}
   \caption{\small{Error rate versus undersampling ratio ($L/N$), access probability ($p$) , antenna number ($M$), access ratio ($p_1/p_2$) and group number ($G$).}}
   \label{err}
   \vspace*{-0.5cm}
\end{figure*}
\subsection{Thresholding Module}
Even after training, there is no guarantee that the proposed auto-encoder can produce an output $\tilde{\boldsymbol{\alpha}} \in \{0,1\}^N$. Thus, it is necessary to design a thresholding module parameterized by threshold $r$ to convert $\tilde{\boldsymbol{\alpha}}$ to the final output of the proposed approach $\hat{\boldsymbol{\alpha}} \in \{0,1\}^N$. We adopt the thresholding module proposed in our previous work \cite{8861085}, and present the details here for completeness. Let $\tilde{\boldsymbol{\alpha}}\in \mathbb{R}^N$ denote the input of the thresholding module. Then, $\hat{\alpha}(n) = \mathbb{I}[\tilde{\alpha}(n)\geq r], n\in \mathcal{N}$. Given $T$ training samples $(\mathbf{x}^{[t]},\boldsymbol{\alpha}^{[t]}),t=1,\cdots,T$, let $P_E(r) \triangleq \frac{1}{T}\sum_{t=1}^T\frac{\|\boldsymbol{\alpha}^{[t]}-\hat{\boldsymbol{\alpha}}^{[t]} \|_1}{N}$ represent the error rate for the given threshold $r$. The optimal threshold $r^*=\mathop{\arg\min}_{r}P_E(r)$ is chosen as the threshold for the hard thresholding module.

\section{Numerical Results}
\label{sim}

In this section, we conduct a numerical experiment on the aforementioned application example. We consider the proposed data-driven approach and five baseline schemes, i.e., the naive data-driven approach based on a deep auto-encoder as illustrated in Section~\ref{approach}, LASSO \cite{6994860}, Group LASSO \cite{qin2013efficient}, AMP \cite{8323218} and ML \cite{8437359}, and evaluate the average error rate of device activity detection $\frac{1}{I} \sum_{i=1}^I\frac{\|\boldsymbol{\alpha}^{(i)}-\hat{\boldsymbol{\alpha}}^{(i)} \|_1}{N}$ and computation time (on the same server) of each scheme over the same set of $I$ testing samples. We choose $N=500$, $\mathbf{h}_m \sim \mathcal{CN}(\mathbf{0}_{N\times 1},\mathbf{I}_{N\times N}), m \in \mathcal{M}$ and $\sigma^2=0.1$. LASSO, GROUP LASSO, AMP and ML use the same set of pilot sequences with the entries generated according to $\mathcal{CN}(0,1)$ in an i.i.d. manner. For the two data-driven approaches, we set $V = 1$, based on a large number of experiments and the tradeoff between performance and computation time.  For a fair comparison, we require ${\|\mathbf{a}_n\|}_2=\sqrt{L}$ in training the architectures of the two data-driven approaches, as in \cite{8861085}. Each data-driven approach adopts the common sensing matrix (pilot sequences) obtained from the encoder of the trained architecture, and uses the decoder of the trained architecture for jointly sparse support recovery (device activity detection). The sizes of training samples and validation samples for training the architectures of the two data-driven approaches and the size of testing samples for evaluating all schemes are $9\times 10^4$, $1\times 10^4$ and $1\times 10^4$, respectively. The training method is the same as that in \cite{8861085}, and is omitted due to page limitation.

To demonstrate how the proposed approach benefits from exploiting properties of sparsity patterns, the following group sparsity model is adopted. Divide $N$ devices into $G$ groups of the same size. The active states of the devices within each group are the same, and there are two group access probabilities, denoted by $p_1$ and $p_2$. Consider $G$ Bernoulli random variables $\xi_j\in{0,1}, j \in \mathcal{G}\triangleq\{1, \cdots, G\}$ with ${\rm Pr}[\xi_j=1]=p_1,j\in \mathcal{G}\cap \{1,3,5\cdots\}$ and ${\rm Pr}[\xi_j=1]=p_2,j\in \mathcal{G}\cap \{2,4,6\cdots\}$. Let $p \triangleq \frac{G_1p_1+G_2p_2}{G}$ denote the average group activity probability, where $G_1\triangleq |\mathcal{G}\cap \{1,3,5\cdots\}|$ and $G_2\triangleq|\mathcal{G}\cap \{2,4,6\cdots\}|$. Note that when $G=N$ and $p_1 = p_2$, device activities become i.i.d.

\begin{figure*}[t]

\begin{center}
 \subfigure[\scriptsize{Computation time versus $L/N$ at $p=0.1$, $M=4$, $p_1/p_2=3$, $G=50$.}]
 {\resizebox{5.2cm}{!}{\includegraphics{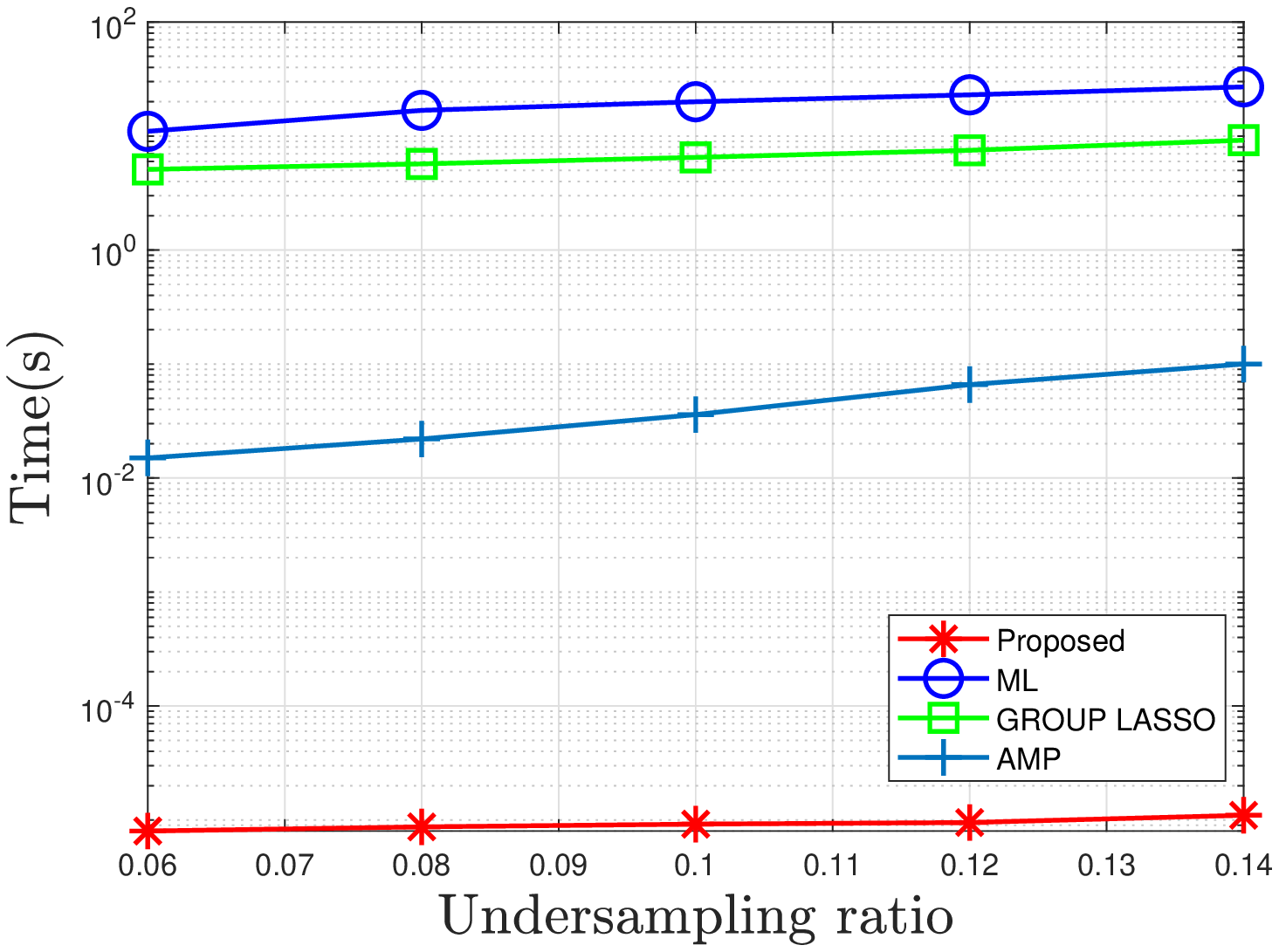}}}
 \subfigure[\scriptsize{Computation time versus $M$ at $L/N=0.14$, $p=0.1$, $p_1/p_2=3$, $G=50$.}]
 {\resizebox{5.2cm}{!}{\includegraphics{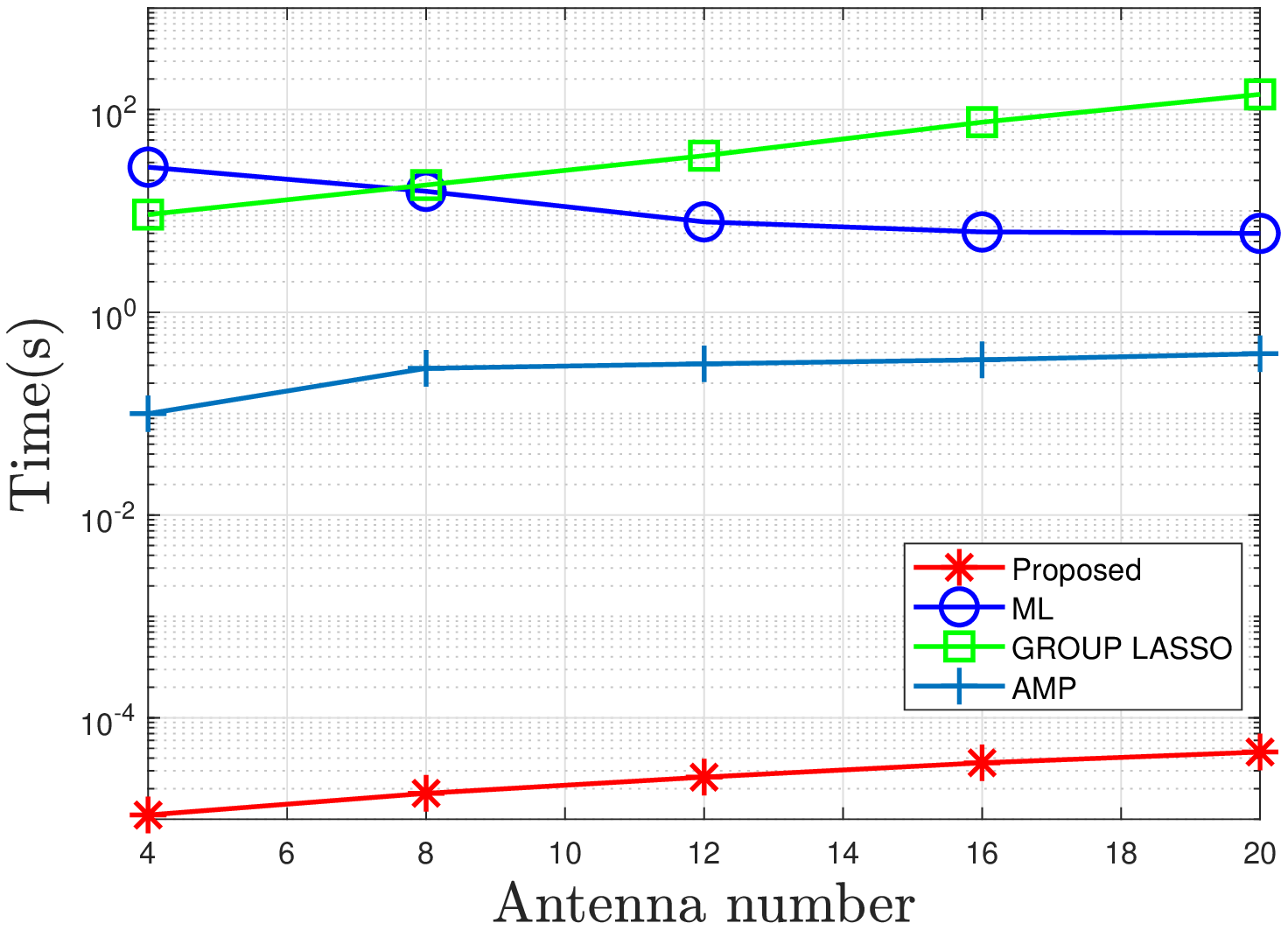}}}
  \end{center}
  \vspace*{-0.4cm}
   \caption{\small{Computation time (sec) versus undersampling ratio ($L/N$) and antenna number ($M$).}}
   \label{time}
   \vspace*{-0.45cm}
\end{figure*}

Fig.~\ref{err} illustrates the error rate versus the undersampling ratio $L/N$, access probability $p$, antenna number $M$, access ratio $p_1/p_2$ and group number $G$.  From Fig.~\ref{err} (a), we can see that LASSO performs much worse than Group LASSO and AMP at small $M$, as explained in Section~\ref{approach}; and the naive approach performs worse than the proposed approach, which demonstrates the benefit of explicitly utilizing the feature of common support in jointly sparse support recovery. Given their unsatisfactory recovery performance, we no longer compare with LASSO and the naive approach in the remaining figures. From Fig.~\ref{err}, we can observe that the proposed approach has the smallest error rate, demonstrating the advantages of the proposed approach in effectively exploring and exploiting sparsity patterns for improving recovery accuracy. From Fig.~\ref{err} (a), (b) and (c), we can see that the error rate of each scheme decreases with $L/N$ and with $M$, and increases with $p$. Fig.~\ref{err}(d) shows that the error rate of each baseline scheme almost does not change with $p_1/p_2$; and the error rate of the proposed approach decreases with $p_1/p_2$, which shows its ability for exploiting the difference in device activity to improve recovery accuracy. The following observations can be made from Fig.~\ref{err}(e). The error rate of Group LASSO seldomly changes with $G$, as $G$ does not affect the optimization problem for Group LASSO. The error rates of ML and AMP both decrease with $G$, as ML and AMP are designed based on the assumption of independent device activity and the device activities become more independent as the group size $N/G$ decreases. The error rate of the proposed approach slightly increases with $G$. The reason is that as $G$ increases, the device activity state space enlarges and it is harder for the neural network to approximate the jointly sparse support recovery process with a fixed number of samples $I$.

Fig.~\ref{time} shows the computation time versus the undersampling rate $L/N$ and antenna number $M$. From Fig.~\ref{time}, we can see that the computation time of the proposed approach is several orders of magnitude lower than those of the baseline schemes, owning to the parallelizable neural network architecture; and AMP has significantly shorter computation time than Group LASSO and ML. Note that the computation time of each scheme depends (almost) only on $N$, $L$ and $M$, and (almost) does not change with the sparsity pattern. In addition, it is worth noting that computation time is an extremely important factor for real-time device activity detection in MIMO-based grant-free random access for mMTC.

\section{Conclusion}
\label{sec:majhead}

In this paper, a data-driven approach is proposed to jointly design the common sensing matrix and jointly sparse support recovery method for complex signals, using a standard deep auto-encoder for real numbers. The proposed approach achieves a substantially lower error rate than classic methods including optimization-based methods, thanks to the effectiveness of the joint design and the ability to exploit structures of sparsity patterns. In addition, the computation time of the proposed method is several orders of magnitude lower than those of the classic methods, owing to the neural network architecture. The proposed approach offers an efficient and effective way for real-time device activity detection in MIMO-based grant-free random access for mMTC.

\bibliographystyle{IEEEtran}
\bibliography{BIBFILE}

\end{document}